\documentclass[prl,amsfonts,twocolumn,nofootinbib,showpacs]{revtex4}
\RequirePackage{graphicx}

\newcommand{\be}{\begin{equation}} 
\newcommand{\ee}{\end{equation}}
\newcommand{\bea}{\begin{eqnarray}}
\newcommand{\eea}{\end{eqnarray}}

\newcommand{\gapp}{\mathrel{\raise.3ex\hbox{$>$}\mkern-14mu
              \lower0.6ex\hbox{$\sim$}}}
\newcommand{\lapp}{\mathrel{\raise.3ex\hbox{$<$}\mkern-14mu
              \lower0.6ex\hbox{$\sim$}}}

\newcommand\lsim{\lesssim}
\newcommand\gsim{\gtrsim}

\renewcommand\({\left(}
\renewcommand\){\right)}
\renewcommand\[{\left[}
\renewcommand\]{\right]}

\newcommand\eq[1]{Eq.~(\ref{#1})}
\newcommand\eqs[2]{Eqs.~(\ref{#1}) and (\ref{#2})}

\newcommand\pa{\partial}

\newcommand\mpl{M_{\rm P}}

\newcommand{\dlabel}[1]{\label{#1}}

\def\calp{{\cal P}}

\def\calpz{{\calp_\zeta}}

\newcommand\bfk{{\mathbf k}}

\newcommand\bfx{{\mathbf x}}

\newcommand\GeV{\,\mbox{GeV}}
\newcommand\MeV{\,\mbox{MeV}}

\newcommand\sub[1]{_{\rm #1}}
\newcommand\su[1]{^{\rm #1}}

\newcommand\mone{^{-1}}
\newcommand\mtwo{^{-2}}
\newcommand\mthree{^{-3}}
\newcommand\mfour{^{-4}}
\newcommand\mfive{^{-5}}
\newcommand\mhalf{^{-1/2}}
\newcommand\half{^{1/2}}

\newcommand\quarter{^{1/4}}

\newcommand{\fnl}{f\sub{NL}}

\newcommand{\calpzphi}{\calp_{\zeta_\phi}}

\newcommand{\calpzsigma}{\calp_{\zeta_\sigma}}

\newcommand{\rad}{\sub {r}}
\newcommand{\sig}{_\sigma}
\newcommand{\dec}{\sub {d}}

\begin{document}


\title{The inflating curvaton}
\author{Konstantinos Dimopoulos}
\affiliation{Department of Physics, Lancaster University, 
Lancaster LA1 4YB, UK}
\author{Kazunori Kohri}
\affiliation{Cosmophysics group, Theory Center, IPNS, KEK,
and The Graduate University for Advanced Study (Sokendai),
Tsukuba 305-0801, Japan}
\author{David H. Lyth}
\affiliation{{}Department of Physics, Lancaster University, 
Lancaster LA1 4YB, UK}
\author{Tomohiro Matsuda}
\affiliation{Department of Physics, Lancaster University,  
Lancaster LA1 4YB, UK, and
 Laboratory of Physics, Saitama Institute of Technology,
Fukaya, Saitama 369-0293, Japan}

\begin{abstract}
The  primordial curvature perturbation $\zeta$ may be generated by 
some curvaton field $\sigma$, which is negligible  during inflation and has
 more or less negligible interactions until it decays. 
In the  current   scenario,  the 
 curvaton  starts to oscillate while its energy density $\rho\sig$ 
is  negligible. We  explore the opposite scenario,  in which 
$\rho\sig$ drives a  few  $e$-folds of inflation before the oscillation
begins. In this scenario for generating $\zeta$
it is exceptionally easy to solve the 
$\eta$ problem; one just has to make the curvaton  a string axion,
with anomaly-mediated susy breaking  which
may soon be tested at the LHC.
The observed spectral index $n$ can be obtained with  a potential
$V\propto \phi^p$ for the first inflation; $p=1$ or $2$ is allowed by the current uncertainty in $n$ but the  improvement in accuracy promised by Planck
may rule out $p=1$.  The predictions include
(i)   running $n'\simeq 0.0026$ ($0.0013$) for $p=1$ (2) 
 that will probably be observed, 
(ii)    non-gaussianity parameter $\fnl\sim -1$ that may be observed,
 (iii)   tensor fraction $r$ is probably too small to ever observed.
\end{abstract}

\preprint{KEK-TH-1501}

\maketitle



{\bf Introduction}---The primordial curvature perturbation $\zeta$
 is already present a
few Hubble times before cosmological scales start to enter
the horizon.
 At that stage, and on those scales, its Fourier components
$\zeta(\bfk)$ are  time-independent  and  set the 
principle (or only ) initial
condition  the subsequent formation of large-scale structure in the Universe
\cite{book}. 
As a result, $\zeta(\bfk)$ can be determined, and 
one of the main tasks of theoretical cosmology is to explore
models of the early universe that can generate it.

The generation of  $\zeta(\bfk)$ 
presumably starts at horizon exit during
inflation ($k=aH\equiv \dot a$ where $a(t)$ is the scale factor of the 
universe) when the vacuum fluctuation of one or more 
scalar (or vector  \cite{dklr})  fields
becomes a  classical  perturbation. According to the  original scenario,
$\zeta$ is generated by   the perturbation $\delta\phi$ of
the  inflaton field  in a single-field
slow-roll inflation model. In that case, $\zeta$ is generated 
promptly at horizon exit, remaining constant thereafter.
According to the curvaton  scenario \cite{ourcurv}, $\zeta$ is instead
generated by the perturbation $\delta \sigma$ of 
 a `curvaton' field, that has practically
no effect during inflation and  generates $\zeta$ 
only when it's energy density becomes a significant fraction of the total.

Instead of the  curvaton scenario one can consider an inflaton-curvaton
scenario where  both $\delta\phi$ and $\delta\sigma$ 
 contribute significantly to $\zeta$ \cite{othercurv}. One can also suppose
 that  $\zeta$ is generated during 
multi-field inflation, or 
by  a `modulating' field that causes an effective mass or coupling
to be inhomogeneous \cite{book}. 
In this Letter we stay with the simpler curvaton scenario.

Up till now, it has been assumed (Figure 1)
  that  the curvaton  starts to oscillate while its energy
density $\rho_\sigma$ is a negligible fraction of the total.
Here we assume instead 
(Figure 2) that $\rho_\sigma$ comes to dominate the total while it is still
slowly varying, giving rise to a second  era  of inflation. 
For simplicity, we
 demand that these  `cosmological scales' are outside  the horizon 
($k<aH$) when the second inflation begins.

A second inflation has been discussed many times before,
but always within a scenario where  the first inflation
generates at least a significant part of $\zeta$, and/or some or all
cosmological scales start out within the horizon.
The  second inflation is usually supposed to begin 
while the  inflaton of the first inflation is still
oscillating.
The contribution of the second inflation for this `double inflation'
is calculated for
instance in \cite{ps,isty} taking cosmological scales to be outside the 
horizon,  and in \cite{anupam2} taking them to be partially inside.
 A  second inflation
starting  during radiation domination  might be called
late inflation. Slow-roll late inflation was considered  in
 \cite{rt},  and fast-roll late inflation in \cite{kl}, 
but these authors ignore the effect on $\zeta$
of the second inflation. There is also thermal inflation
\cite{thermal} which really has no effect on $\zeta$. Our scenario
is different from all of these.

\begin{figure}[htb]
\centering
\includegraphics[width=0.8\columnwidth]{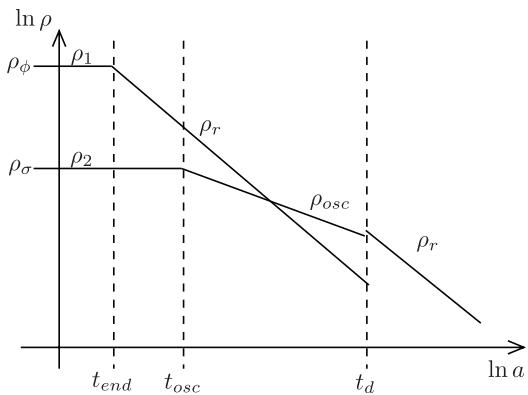}
\caption{Original curvaton scenario. The inflaton $\phi$
gives nearly constant energy density  $\rho_\phi$ during inflation
which after reheating (shown as instantaneous) converts to radiation
with $\rho\rad\propto a\mfour$.
The curvaton density $\rho\sig$ varies slowly with  $\rho\sig\ll \rho\rad$, 
until $\sigma$ starts to oscillate giving $\rho\sig \propto a\mthree$.
It is assumed that $\zeta$ is constant after the curvaton decays,
which is guaranteed if the universe is then radiation dominated
as in the Figure.}
\label{fig:osc}
\end{figure}

\begin{figure}[htb]
\centering
\includegraphics[width=0.8\columnwidth]{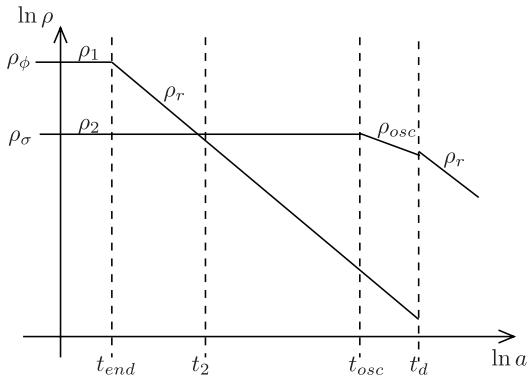}
 \caption{Inflating curvaton scenario. Here $\rho\sig$ 
 exceeds $\rho\rad$ before $\sigma$ oscillates,  giving  a few
$e$-folds of inflation before the oscillation starts.}
\label{fig:inf}
\end{figure}

{\bf Duration of the second inflation}---By `cosmological' scales we mean
those  probed more or less directly by observation.
They  range from $k=a_0H_0$ where 
$0$ denotes the present, 
 to around $k=e^{15}a_0H_0$. 
According to a standard calculation \cite{book}, 
requiring that they are outside the horizon when the second inflation starts
corresponds to 
\be
N_2    \lsim  
 45 -  \ln(10\mfive \mpl/H_2)/2
, \dlabel{n2} \ee
where $N_2\equiv \Delta (\ln a)$ is the $e$-folds of expansion during the
second inflation, $H_2\equiv \dot a/a$ is the Hubble parameter then
and  $\mpl\equiv (8\pi G)\mhalf=2\times 10^{18}\GeV$.
[{\em A subscript} 1(2) {\em will always denote the first (second) inflation}.]

The  right hand side of \eq{n2} takes $\rho=3\mpl^2 H^2$
to be  constant during the second inflation and then $\propto a\mfour$
(radiation) till the observed matter-dominated era. Replacing some of the
radiation domination by matter domination reduces it.
The second term is positive because $H_2<H_1$ and we need
$H_1< 10\mfive \mpl$ or the tensor perturbation with spectrum
$\calp_h= (8/\mpl^2)(H_1/2\pi)^2$ would have been observed
(see \eq{calpzphi} below).

CDM and baryon number cannot be created
before the curvaton (or any other mechanism) creates $\zeta$,
since that would give 
an  isocurvature perturbation excluded by observation  \cite{ourcurv}.
That requires something like $\rho_2\quarter>10^3\GeV$ 
corresponding to $H_2>10^{-30}\mpl$ 
and $N_2 < 16$. {\em We therefore
require roughly $N_2< 45$ to $16$}.

{\bf Calculating the curvature perturbation}---The 
curvature perturbation $\zeta$ is  described  non-perturbatively through the
$\delta N$ formalism as in  \cite{lr}.
In this paper, we  just  work to first order in $\delta\rho$,
as  in \cite{ourcurv}. Then
\begin{equation}
\zeta(\bfk,t)=-H(t)\frac{\delta\rho(\bfk,t)}{\dot\rho(t)} 
= \frac13 \frac{\delta\rho(\bfk,t)}{\rho(t)+p(t)},
\dlabel{zetadef}
\end{equation}
where  $p$ is the pressure, and 
$\delta\rho$ is defined on the slicing with  uniform locally-defined scale factor
$a(\bfx,t)$
(flat slicing). The second equality corresponds to
the energy continuity equation  $\dot\rho=-3H(\rho+p)$. 

Keeping only super-horizon scales, the 
 energy continuity equation is valid locally. As a result 
\cite{constantz},  $\zeta$  is constant during any era when  
 $p(\bfx,t)$  is a unique function of $\rho(\bfx,t)$; that is the case for
 pure   radiation
 ($p=\rho/3$) or  matter ($p=0$).
When cosmological scales start to enter the horizon, the temperature
$T$ is somewhat less than $1\MeV$ and 
we know that the Universe is practically pure radiation
giving a time-independent  $\zeta(\bfk,t)$ that we are denoting simply
by $\zeta(\bfk)$. 
According to the  
curvaton scenario, $\zeta(\bfk,t)$ does not vary between 
 curvaton decay at $t=t\dec$ and $T\sim \MeV$, which is  guaranteed if
  the universe is  completely radiation-dominated throughout
that era.

In any curvaton scenario,  $\zeta$ is generated while  $\rho=\rho\sig
+\rho\rad$ and $p=p\sig+p\rad$, where 
$\rho\sig = V(\sigma) + \dot\sigma^2/2$ and
$p\sig = -V(\sigma) + \dot\sigma^2/2$ are the curvaton contributions.
 For the original
curvaton $\rho\rad$ cannot be matter and is taken to be pure 
radiation, $\rho\rad\propto a\mfour$
 {\em For the inflating curvaton, $\rho\rad$ might be matter,
$\rho\rad\propto a\mthree$
 allowing it to correspond to the oscillation of a  field
$\phi$ responsible for the first inflation.} It might even 
decrease more slowly, say like $a\mone$ corresponding to a cosmic
string network \cite{anupam}. In these cases though,
we demand that
$\rho\rad$ becomes radiation before it is
 a significant fraction
of $\rho$, so that it does not  cause  $\zeta$ to vary significantly.

To facilitate an analytic calculation, one 
 writes \eq{zetadef} as $\zeta(\bfk,t) = f(t) \zeta\sig(\bfk,t)$
where $3\zeta\sig\equiv (\delta\rho\sig)/(\rho\sig+p\sig)$
and $f(t) \equiv (\rho\sig + p\sig)/(\rho+p)$. 
There is supposed to be  negligible exchange of energy between 
the two components, so that 
 $\zeta\sig$ is constant  if 
 $p\sig(\bfx,t)$  is a unique function of $\rho\sig(\bfx,t)$.
 For the original curvaton,
$\zeta\sig$ becomes constant only after the oscillation begins, when
$p\sig\simeq 0$. We now argue that for the inflating curvaton,
 $\zeta\sig$ will become  constant soon after the second inflation begins
at the epoch $t=t_2$. We begin with the following equation, valid
 in the absence of perturbations \cite{book}:
\be
\ddot \sigma  + 3H\dot \sigma  + V'(\sigma) =0  \dlabel{sigmaeq}.
\ee
Since sub-horizon modes of $\sigma(\bfx,t)$ redshift away 
that quantity has negligible spatial gradient \cite{book}.
As a result it satisfies \eq{sigmaeq} for each $\bfx$, 
 with $3\mpl^2H^2
=\rho\sig+\rho\rad$  the locally defined
quantity and  $t$ the  proper time. Also, 
 $p\sig$ is a unique function of $\rho\sig$ if and only if
$\sigma(\bfx,t)$ is unique up to the choice of $t=0$
(attractor solution).
We expect that soon after the inflation begins it will be slow-roll
or fast-roll (see below), and both of these make $\dot\sigma$
 a unique function of $\sigma$ giving indeed the required attractor
solution. 

If $\rho\rad$ were  completely negligible  we could  invoke a more
general argument for the attractor \cite{bs}, but that might not 
apply because
although the contribution of $\rho\rad$
to $H$ becomes small soon after the second inflation begins,
its contribution to
\be  
2\epsilon_H \equiv 2|\dot H|/H^2
=3\(\dot\sigma^2 + \frac43  \rho\rad \)/\rho
\dlabel{eph2} \ee
may be  dominant at least initially. 
While that is happening,
$f(t) \simeq (\dot\sigma^2)/(4\rho\rad/3)\ll 1$. But 
just before $\sigma$ decays at $t\dec$ we have
$f(t\dec)\simeq \rho\sig/[\rho\sig + (4/3)\rho\rad]$ which
will be very close to 1,
{\em In contrast, the oscillating curvaton can have $f(t\dec)\ll 1$.}

 Keeping only super-horizon modes, $f(t\dec)=1$ gives 
 $\zeta(\bfx) = \delta\rho\sig(\bfx,t_2)/3\dot\sigma^2(t_2)$. To 
 first order in $\delta\sigma$,
$\zeta(\bfx) =   V'\delta\sigma(\bfx)/3\dot\sigma^2$.
At horizon exit during the first inflation $\delta\sigma$
is nearly gaussian with spectrum $H_1/2\pi$. Allowing 
$\sigma(\bfx,t_2)$ to be a function $g$ of its value at horizon
exit, but taking both $\sigma(\bfx,t)$ and  $H_1$ to be time-independent while
cosmological scales leave the horizon, we  get  the  scale-independent 
spectrum
\begin{equation}
\calp\half_\zeta  \simeq  \frac{g'}{3} \frac{V'(\sigma(t_2))}
{\dot\sigma_2^2(t_2)}  
\frac{H_1}{2\pi},
\end{equation}
Observation gives \cite{wmap7}
$\calp_\zeta\half \simeq  5\times 10\mfive$.

Taking into 
account the time-dependence of $H_1$ and $\sigma$
while cosmological scales leave the horizon one finds \cite{book}
\be
n(k)-1\equiv d\ln \calp/d\ln k = -2\epsilon_{H1} + 2 \eta_1
\dlabel{neq}, \ee
with the right hand side evaluated at horizon exit $k=aH$
and $3H^2\eta \equiv \pa^2 V/\pa \sigma^2$.  
[Note that $V =V(\sigma,\phi,\cdots)$ during the first inflation
where $\phi,\cdots$ are the fields responsible for that  inflation.]
Assuming a tensor fraction $r\equiv \calp_h/\calpz \ll 10\mone$ and 
$|n'|\ll 10\mone$ where $n'\equiv dn/\ln k$,
 observation \cite{wmap7} 
 gives $n-1=-0.037\pm0.014$.

\eq{neq}  is a universal formula, applying whenever a field $\sigma$ different
from the inflaton generates $\zeta$.  
It can easily happen  (as in our case, see below)
 that the last term is negligible. Then  we need 
$2\epsilon_{H1}\simeq    0.037$, leading to three important consequences.

1. To get the required  $\epsilon_{H1}$ 
we need  \cite{mytensor}
a large   change in the  inflaton field $\phi$ 
during the first inflation. 
To achieve that one usually takes $\phi$ to have the canonical kinetic term
with $V(\phi,\sigma) \simeq V(\phi) = A \phi^p$. 
Then  $\epsilon_{H1} \simeq p/4N(k)$, 
where $N(k)$ is the number of $e$-folds of 
the first inflation after the scale $k$ leaves the horizon \cite{book}.
Defining  $N_1\equiv N(a_0H_0)$, we  
 need $N_1\simeq 14p$ to get the required
$2\epsilon_{H1}\simeq    0.037$.
 While one is free to postulate any $p$ 
{\em 
the only choices of} $p$ {\em 
with good justification are} $p=1$ {\em (corresponding to
monodromy} \cite{mono}) 
{\em and} $p=2$ ({\em corresponding to `extranatural'} \cite{extra}
{\em inflation or} $N$-{\em flation} \cite{nflation}. 
{\em These give $N_1\simeq 14$ and $28$ respectively}.

2. A standard calculation   \cite{book} gives
$N_1  \simeq 60    -   \ln(10\mfive\mpl/H_1)/2  - N_2$,
where the equality would be exact if $\rho$ were constant during both
inflations and $\propto a\mfour$ otherwise until the observed matter-dominated
era. Combining this with \eq{n2} gives
\be
15 + \frac12\ln(H_1/H_2) \lsim N_1 \lsim 60 - \frac12\ln(10\mfive\mpl/H_1) 
. \dlabel{n1eq} \ee
Taking into account the uncertainty in $n$ {\em this is 
compatible with} $N\simeq 14p$ {\em for} $p=1$ {\em or} $2$,
{\em though} $p=1$ {\em
may be ruled out when Planck reduces the uncertainty.
For the oscillating curvaton, 
where} $N_2$ {\em is absent, we would probably need}  $p\sim 3$ {\em to} $4$. 

3. Since  $n(k)-1\propto 1/N(k)$ the `running' $n'$ is given by
$n'=(1-n)/N_1$. This prediction holds also for the oscillating curvaton
scenario (if $|\eta_1|\ll \epsilon_{H1}$), 
and for the inflaton scenario within some simple slow-roll
models and it  makes $n'$  big enough to observe in the future. 
For the oscillating curvaton,  and the  inflaton scenario, 
 one expects roughly $N_1\simeq 60$ corresponding to
$n'\simeq 0.0007$. For the inflating curvaton, we have for
$V(\phi)\propto \phi^p$ and 
taking account of the  uncertainty in $n$
\be
n' = \( \frac{1-n}{0.037} \)^2 \frac{0.0026}p \dlabel{running}
, \ee
with $p=1$ or $2$.
{\em This precise prediction for $n'$ will be probably be tested in the future
(see for instance} \cite{nmeas}).

We require the contribution
of the first inflation to be negligible, 
$s\equiv \calpzphi/\calpz \ll 1$. Assuming canonical kinetic terms
for the inflaton(s) \cite{book},
\be
\calpzphi\half \geq \frac1{ \sqrt{2\epsilon_{H1}} }\frac{H_1}{2\pi\mpl}
\dlabel{calpzphi}  \ee
and   $r\leq   16s\epsilon_{H1}$ implying $H_1\leq 1.1\times 10\mfour
(rs)\half\mpl$. 
(The equalities  hold for a single inflaton.) Observation \cite{wmap7}
gives $r\lsim 10\mone$.
Even if the second  term of \eq{neq} contributes significantly, it is
 unlikely to accurately cancel the first term which means that
 we need $\epsilon_{H1}\lsim 0.02$, giving
 $r\lsim 0.3s$ and $H_1\lsim 6\times 10\mfive s\half\mpl$.
The tensor fraction will  not be observed 
by Planck \cite{planck} if $s\lsim 10\mone$,
and it will probably never be observed \cite{kazref} if
$s\lsim 10\mthree$.

To calculate the non-gaussianity parameter  $\fnl$ we 
expand $\delta\rho\sig$ to second order in $\delta\sigma$ giving
\bea
\zeta(\bfx) &=& V'\delta\sigma/3\dot\sigma^2 + (3/5) 
\fnl (V'\delta\sigma/3\dot\sigma)^2
\dlabel{zetasecond}  \\
\fnl &=& \( 5 \dot\sigma^2 V''/V'^2 \) \( 1+ g''/g'^2 \)
. \dlabel{fnl}\eea
Observation \cite{wmap7}
requires  $-10<\fnl<74$ which means that the second term of
\eq{zetasecond} gives a negligible contribution \cite{bl} to $\calpz$.
We  need $|\fnl|\gsim 1$ if $\fnl$ is ever to be detected.
With such a value
it will indeed  
 be a good approximation to ignore the non-gaussianity of $\delta\sigma$
\cite{vw}.
But our first-order treatment of $\delta\rho\sig$ is reliable 
only \cite{book} for $|\fnl|\gg 1$; for $|\fnl|\sim 1$
 one should  go to second order, or  use the $\delta N$ formalism
as in \cite{ourcurv}.

The usual way of achieving inflation would be through 
the slow roll approximation $\dot\sigma\simeq - V'/3H_2$. 
That 
will typically make $N_2$ too large but let us anyway see what it implies.
  Differentiating it requires 
 $\epsilon_{H2}
\ll 1$ and $|\eta_2| \ll 1$ where 
 $\eta_2 \equiv V''/3H_2^2$. Using \eq{eph2} the former condition 
requires $\epsilon_2 \equiv \mpl^2 (V'/V)^2/2 \ll 1$ and 
$\rho\rad \ll \rho$.  Then we get 
\bea
\calp\half_\zeta&=& (2\epsilon_2)\mhalf  (H_1/2\pi) (g'/\mpl) \\
(3/5)\fnl&=& \eta_2 \( 1 + g''/g'^2 \)
. \dlabel{fnl2} \eea
We  would need $|g''|\gg g'^2$ to get a detectable $\fnl$.

Most discussions of the oscillating 
curvaton take $V(\sigma)\simeq m_\sigma^2 \sigma^2/2$.
{\em This choice is impossible for the inflating curvaton
scenario}
if the  first inflation has inflaton(s) with canonical
kinetic terms. 
 Indeed, using \eqs{calpzphi}{fnl2} 
 the curvaton contribution $\calpzsigma$ is given by
$\calpzsigma/\calpzphi =2\epsilon_{H1} N_2(\sigma_2/\sigma_1)^2\lsim 1$
which means that $\calpzsigma$ cannot  dominate.

{\bf The curvaton a string axion}---Any scheme for generating $\zeta$ 
from some   field $\chi$ encounters
the $\eta$ problem; that a generic supergravity theory 
gives \cite{lm}
in the early universe $|\eta_\chi\su{nr}|\gsim 1$,
where $\eta_\chi\su{nr} \equiv (\pa^2 V/\pa \chi^2)/3H\sub{nr}^2$ 
and $3\mpl^2H^2\sub{nr} \equiv \rho\sub{nr}$ excludes any  
radiation contribution.  
It is a problem for two reasons; (i)  we need $|\eta_\chi|\lsim 
10\mtwo$ while cosmological scales leave the horizon to keep
$|n-1|$ small enough, (ii) unless $|\pa V/\pa \chi|$
is exceptionally small we generally need  $|\eta_\chi|\ll 1$ at all times
or $\chi$ will  be quickly driven to a minimum of $V$. 

For the inflaton scenario, the  $\eta$ problem exists
only during inflation when $H\simeq$\,const; it is often ignored and
could  be  regarded as a 
fine-tuning requirement on the parameters of the supergravity inflaton
potential.
For the oscillating curvaton scenario 
the $\eta$ problem  may be  more severe because it may exist for a long
time after inflation, with $H$ strongly decreasing.
For the inflating curvaton it  is definitely more severe because it exists
during both inflations with very different values for $H$.

To avoid the $\eta$ problem for the curvaton one can  take
it be a pNGB with the  potential 
\be
V(\sigma) =  \frac12 V_0 \[ 1+ \cos\(\frac{\pi \sigma}{\sigma_0}\)  \] 
, \dlabel{pngb} \ee
practically independent of other field values.
It is known \cite{ourcurv} that  $\sigma_0\ll\mpl$ gives 
 the  oscillating curvaton,  but it is unclear how to motivate
such a value.  Choosing instead $\sigma_0\gg\mpl$ would give
 a second inflation with $|\eta_2|\ll 1$, 
but $N_2$ would typically be too big and this
choice anyhow seems impossible within string theory \cite{michael}.
(If  the latter difficulty is ignored we can replace
$\sigma$ in \eq{pngb} by $\phi$ to arrive at `Natural Inflation' 
\cite{book}.)

What we need for the inflating curvaton is 
 $\sigma_0\sim \mpl$. Then, 
 in the regime $\sigma\ll \sigma_0$ we have
$V\simeq V_0-m^2\sigma^2/2$ giving
 $|\eta_2|\sim 1$  and just a few $e$-folds
of inflation.
Setting $3\mpl^2H_2^2=V_0$ gives
$H_2^2/m^2  \simeq (2/3\pi^2) (\sigma_0^2/\mpl^2)$.
{\em The  required value} $\sigma_0\sim \mpl$ {\em  is achieved if} 
$\sigma$ {\em is a string axion with 
 gravity- or anomaly mediated susy breaking.}
\cite{michael}.
Then  $m$ is  of order the
gravitino mass $m_g$ \cite{michael}. 
The curvaton and gravitino have to decay  before they can upset BBN 
 which requires \cite{kaz} $m\gsim 10^4\GeV\sim 10^{-14}\mpl$.
{\em This corresponds to   anomaly mediation,  which gives}
$m\sim 10^4$  {\em to}  $10^5\GeV$ {\em and (like any
 version of susy)   may soon be tested at the} LHC.

The contribution of $\eta_1$ to $n-1$ will be negligible if
$H_1/m\gg [3(1-n)/2]\mhalf =4$. With such a low $m$ this is a mild requirement
which we will take to be satisfied, so that \eq{running} holds.

Writing   $V'=-m^2\sigma$, and taking $H_2$ to be constant,
 \eq{sigmaeq}  gives during the second inflation 
\begin{equation}
\dot\sigma \simeq  FH_2 \sigma,\qquad
F \equiv - \frac32 + \sqrt{\frac94 + \frac{m^2}{H_2^2} } 
\sim m/H_2
\dlabel{fast}
. \end{equation}
The  slow-roll regime is $m\ll H_2 $ but we are interested in the 
`fast-roll' \cite{fastroll} regime
  $m\gsim H_2$ corresponding to 
$F\gsim 1$. (The approximation $F\simeq  m/H_2$ is adequate for
 $m^2/H_2^2\gsim 3$.)
In the fast-roll  regime,
\eq{fast} is self-consistent if $N_2\gg 1$ and $\sigma\ll \sigma_0$.

Since \eq{sigmaeq} is linear, $g$ is a linear function and we get
\bea
\calp\half_\zeta  &=& \frac13 \( \frac{m}{FH_2} \)^2 \frac{H_1}{2\pi\sigma_1}
\sim  \frac{H_1}{6\pi\sigma_1} \dlabel{calpfr}\\
(3/5)\fnl &=& - (FH_2/m)^2  \sim    -1  \dlabel{fnlfr}
. \eea
The result for $\fnl$ may be 
strongly modified by the correction of second order
in $\delta\rho_\sigma$, but barring a strong cancellation {\em it seems that
$\fnl$ may eventually be detectable}.

Since inflation ends at $\sigma\sim \sigma_0$ we have 
  $FN_2 \sim \ln (\mpl/\sigma_2)$.
Since $N_2\gg 1$,   we have   $F\ll FN_2$, hence 
$m/H_2\ll FN_2$.
 Using \eq{calpfr} with $ H_1>H_2\gsim 10^4\GeV$
gives  $FN_2\lsim 24$.
Going the other way,   \eq{calpfr} with $H_1\sim  10^{-6}\mpl$ implies
$FN_2\lsim 4$.

\eqs{calpfr}{fnlfr} 
 are roughly the same as those of 
the oscillating  curvaton model with 
$f(t\dec)\simeq 1$.  But the result for
 $n'$ is different for the oscillating curvaton \cite{book}; there we might
have  $n-1\simeq 2\eta_1$ with $n'$ negligible, and even if 
\eq{running} holds we expect $p\simeq 4$.

One may worry about the 
assumption that $\sigma(t_1)$ is  near the top of the potential,
given that the first inflation may be of long  duration. 
For a given $H_1(t)$, the late-time probability 
distribution of $\sigma$ at the end of the first inflation can be
calculated \cite{stochastic}. Taking that distribution to apply and also 
taking $H_1$ to be constant,  we would need
$H_1^4 \gsim V_0$ to have a significant probability that  $\sigma$ is
near the top. This requires
 $H_1/H_2 \gsim \mpl/H_1 \gsim 10^5$ and (since 
   $H_2> 10^4\GeV$)  $H_1 \gsim 10^{-7}\mpl$. The former  bound
would probably make  \eq{n1eq} incompatible with $p=1$. But to know whether  the
estimate $H_1^4 \gsim V_0$ is realistic one would have to calculate 
the evolution of the 
probability distribution with the correct $H_1(t)$ and a range of initial values of $\sigma$.

{\bf Conclusion}---The hypothesis that  the curvaton is a string axion
leads to a simple early-universe scenario.
The  curvaton generates a few $e$-folds
of inflation with $H\sim 10^4\GeV$, 
during which $\zeta$ is created. 
The main inflation takes place earlier,
with a potential $V(\phi)\propto \phi^p$, with $p=1$ or 2 needed to reproduce
the observed spectral index within current observational uncertainity.
The hypothesis  requires  low-energy
susy with anomaly-mediated susy breaking which may soon be tested at the
LHC. It predicts that a tensor fraction
$r$ that is probably too small ever to observe, but 
a   running $n'$  that  eventually be observed and will
 decide between the linear and quadratic potentials.
A  third  prediction $\fnl\sim -1$ may also be testable, but the accuracy
of the calculation needs to be improved.

TM and KK thank Anupam Mazumdar for valuable discussions in the 
early stage of this work. DHL thanks Michael Dine for correspondence, and
has  support from
 UNILHC23792, European Research and Training Network (RTN) grant.
KK is partly supported by the Grant-in-Aid for the Ministry of  
Education, Culture, Sports, Science and Technology, Government of
Japan Nos. 21111006, 22244030, 23540327, and by the Center for the 
Promotion of Integrated Science (CPIS) of Sokendai (1HB5806020).

\end{document}